\documentclass[%
 reprint,
 amsmath,amssymb,
 aps,
]{revtex4-2}

\usepackage{graphicx}
\usepackage{dcolumn}
\usepackage{bm}
\usepackage{hyperref}
\usepackage{float}

\begin{document}

\preprint{APS/123-QED}

\title{Resolving Microscopic Correlated Electron Dynamics via 2000-Qubit Quantum Simulation}

\author{Jaka Vodeb}
\email{Corresponding author: Jaka Vodeb, email: jaka.vodeb@ijs.si}
\affiliation{%
 Department of Complex Matter, Jožef Stefan Institute, Jamova 39, 1000 Ljubljana, Slovenia\\
 CENN Nanocenter, Jamova 39, 1000 Ljubljana, Slovenia
}%




\date{\today}

\begin{abstract}
Understanding how quantum materials return to equilibrium after being driven into excited states is a fundamental problem in condensed matter physics. A prototypical material, 1T-TaS$_2$, exhibits complex electronic textures made up of domain walls, which slowly reorganize into a more uniform structure as the system relaxes. At low temperatures, this process becomes dominated by quantum rather than thermal effects. In this work, we use large-scale noise-driven quantum simulations—spanning more than 2000 qubits—to study this relaxation process through an effective model known as the transverse-field Ising model in a longitudinal field. By mathematically transforming this model into a simpler form, we identify the basic microscopic steps involved: rather than moving collectively, the domain walls evolve through a sequence of noise-driven single-particle tunneling events. A detailed analysis of how the relaxation rate depends on temperature and model parameters confirms this picture. Our findings show that quantum simulation can provide rare, predictive insight into the inner workings of real quantum materials, and establish a practical pathway for studying complex non-equilibrium processes using current-generation quantum hardware.
\end{abstract}

\maketitle



\section{Introduction}

Quantum simulation has become a powerful tool for probing many-body systems whose microscopic dynamics are inaccessible to classical computation or direct experiment.
Recent progress across diverse platforms has enabled controlled studies of quantum chemistry~\cite{kandala2017hardware,cao2019quantum}, quantum criticality~\cite{greiner2002quantum,king2023quantum}, lattice gauge theories~\cite{dalmonte2016lattice,vodeb2025stirring}, and disorder-driven localization~\cite{schreiber2015observation}, demonstrating that programmable quantum devices can act as microscopes for complex quantum dynamics.
An emerging challenge is to leverage these capabilities to resolve non-equilibrium relaxation processes in strongly correlated materials, where interactions, disorder, lattice frustration, and topology intertwine.

The layered dichalcogenide 1T-TaS$_2$ provides a paradigmatic example.
At low temperatures it forms a commensurate charge-density wave (CDW) composed of localized polarons arranged in a triangular superlattice~\cite{fazekas1979electrical,brazovskii2015modeling,karpov2018modeling,gerasimenko2019quantum}.
Scanning tunnelling microscopy (STM) reveals extended networks of narrow domain walls separating distinct CDW domains~\cite{vodeb2024non}.
Time-resolved STM has shown that these domain-wall textures relax extraordinarily slowly—over hours—through discrete reconfiguration events, despite underlying electronic energy scales being orders of magnitude faster (femtoseconds)~\cite{vodeb2024non}.
At sufficiently low temperature, the relaxation becomes temperature independent, suggesting a crossover from thermally activated to noise-assisted quantum dynamics~\cite{vodeb2024non}.
Identifying the microscopic mechanism underlying this regime remains an open problem.

In previous work, we demonstrated that a transverse-field Ising model (TFIM) with a longitudinal field, implemented on a superconducting quantum annealer, faithfully reproduces both the spatial structure and the relaxation phenomenology of CDW domain walls in 1T-TaS$_2$~\cite{vodeb2024non}.
While this established a quantitative correspondence between experiment and quantum simulation, it left unresolved two central questions: why is the timescale so slow, and does domain-wall motion arise from coherent collective tunnelling of many particles, or from sequences of local reconfiguration events?

In this work, we resolve this question by combining microscopic modeling, controlled quantum simulation, and dynamical scaling analysis.
Starting from a low-temperature electron model of 1T-TaS$_2$, we make the connection to the TFIM, amenable to quantum simulation, and then derive an effective Hamiltonian of the TFIM in the low-energy sector using a Schrieffer--Wolff transformation~\cite{bravyi2011schrieffer}, showing that domain-wall motion in our quantum simulations is governed by second-order single-particle tunnelling processes.
The intrinsic tunnelling processes occur on timescales much shorter than the observed reconfiguration dynamics and therefore do not set the relaxation rate; instead, the rate is controlled by slow, quasi-static fluctuations of local biases that intermittently enable tunnelling.

This picture is supported by a scaling collapse of the reconfiguration rate as a function of effective temperature and transverse field, revealing only a weak dependence on the tunnelling control parameter.
The absence of the expected strong tunnelling scaling provides direct evidence that domain-wall relaxation is noise assisted rather than tunnelling limited.
Despite differences in global charge evolution between experiment and simulation, the elementary local processes governing domain-wall motion are identical in both systems.

Our results establish a microscopic, noise-driven mechanism for domain-wall relaxation in 1T-TaS$_2$ and demonstrate how programmable quantum simulators can be used to disentangle intrinsic quantum processes from environmental effects in strongly correlated materials.
More broadly, this work highlights quantum simulation as a powerful approach for resolving non-equilibrium relaxation pathways in complex quantum matter.


\section{Results}


\subsection{Low-temperature electronic model and reduction to an effective domain-wall description}

\begin{figure}[t]
    \centering
    \includegraphics[width=\linewidth]{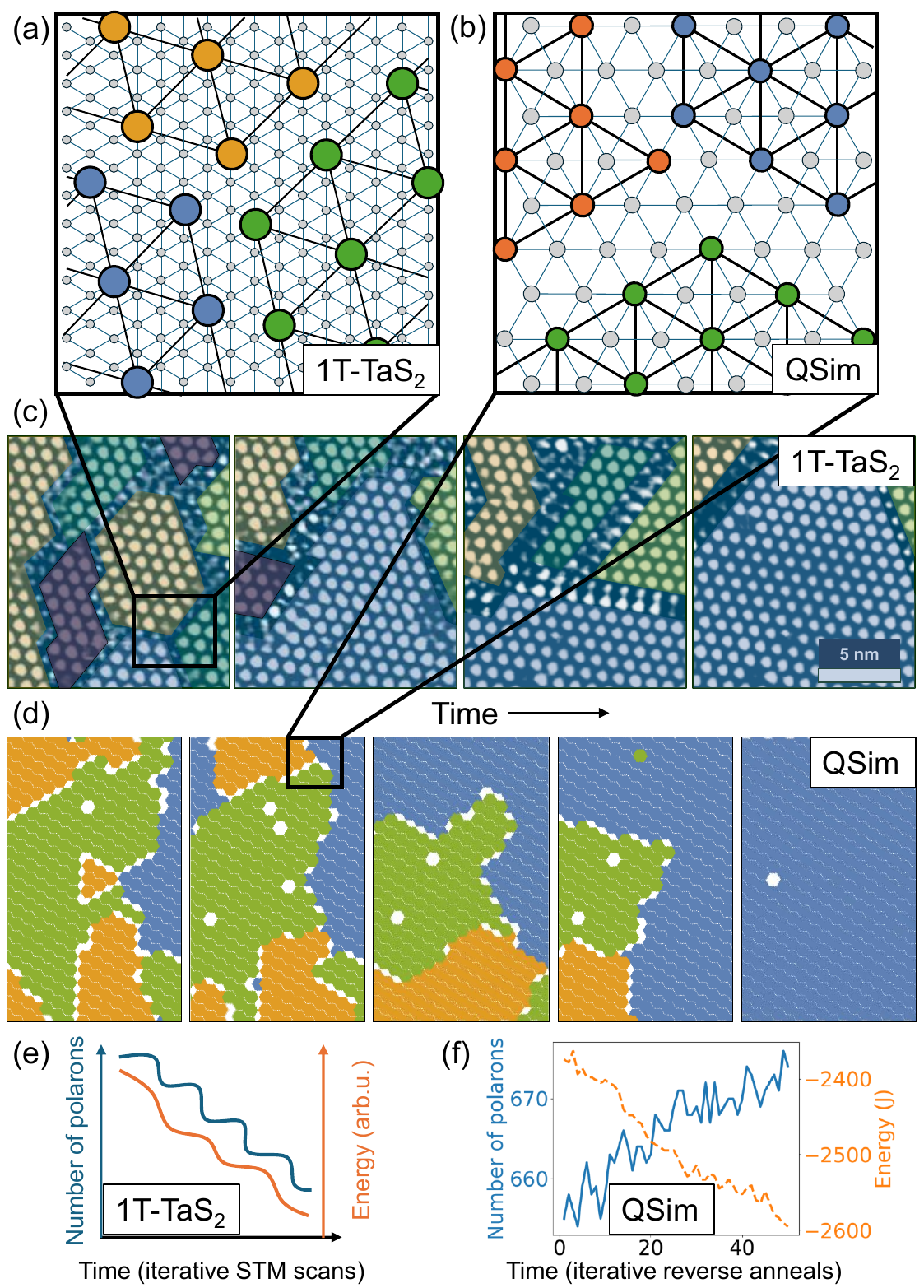}
    \caption{
    \textbf{Domain structures and relaxation in 1T-TaS$_2$ and in quantum simulation.}
    (a)~Representative CDW domain configuration in 1T-TaS$_2$, where colors denote the thirteen distinct commensurate CDW domains related by relative translations of the $1/13$-filled polaronic superlattice.
    Extended domain walls separate regions of different CDW domains and store excess interaction energy.
    (b)~Corresponding domain configuration obtained in the quantum simulation on a quantum annealer.
    Colors label the three degenerate domains of the effective triangular-lattice transverse-field Ising model.
    Despite the reduced number of domain types, the local domain-wall geometry is the same as in the material.
    Black outlines and lines show (a) and (b) as insets of (c) and (d).
    (c)~Time-resolved STM images of 1T-TaS$_2$ showing slow relaxation of an initially non-equilibrium domain-wall network toward the ordered ground state.
    (d)~Representative sequence of configurations from the quantum simulation obtained via iterative annealing cycles, showing domain-wall annihilation and coarsening.
    (e)~Evolution of polaron density and interaction energy extracted from STM measurements during relaxation in 1T-TaS$_2$, where domains are typically offset closer together due to an excess of charge and gradually reconfigure into a single domain through charge leakage.
    (f)~Evolution of polaron density and interaction energy obtained from the quantum simulation.
    The system initializes with missing charge, evidenced by a typical neighboring domain offset moving them away from each other, and leading to an increase in polaron number during relaxation.
    Despite this global difference, the local polaronic dynamics within domain walls are the same in experiment and simulation, as shown in this work.
    }
    \label{fig:1}
\end{figure}

At low temperatures, 1T-TaS$_2$ forms a commensurate charge-density wave (CDW) that can be viewed as a $1/13$-filled triangular Wigner crystal of localized electrons (polarons), corresponding to one electron per thirteen Ta atoms
\cite{fazekas1979electrical,brazovskii2015modeling,karpov2018modeling,vodeb2019configurational,gerasimenko2019quantum}.
The uniform CDW configuration constitutes the ground state (Fig.~\ref{fig:1}a), while STM experiments routinely observe extended networks of narrow domain walls separating regions with different CDW registries (Fig.~\ref{fig:1}c)~\cite{vodeb2024non}.
Time-resolved STM further reveals that these domain-wall networks relax extremely slowly—over hours—via discrete reconfiguration events, despite electronic timescales being measured in femtoseconds.
Remarkably similar relaxation dynamics are observed when analogous non-equilibrium domain configurations are encoded and evolved on a quantum annealer (Fig.~\ref{fig:1}b,d).

Although the global evolution differs—polaron density decreases in experiment but increases in simulation (Fig.~\ref{fig:1}e,f)—the domain-wall structure is closely similar in both cases. Different global dynamics occur simply due to different initial states—electron-doped in 1T-TaS$_2$, and hole-doped in our quantum simulation.
Domains in both cases differ by atomic lattice translations, producing narrow domain walls that govern the slow dynamics.
In both systems, relaxation proceeds through a reduction of interaction and chemical-potential energy, motivating a focus on domain-wall reconfiguration rather than global charge transport.

To model this behavior, we start from a microscopic Hamiltonian for interacting electrons on the triangular Ta lattice, including hopping, long-range Coulomb repulsion, and electron–phonon coupling,
\begin{equation}
\begin{aligned}
\hat H
&=
\sum_{i,j,s,s'}
\left(
t_{i,j} - \mu \delta_{i,j}
\right)
\delta_{s,s'}
\, \hat c_{i,s}^\dagger \hat c_{j,s'}
\\
&\quad
+
\sum_{q,i,s}
\omega_q \hat n_{i,s}
\left(
u_{i,s}(q)\hat d_q + \mathrm{h.c.}
\right)
\\
&\quad
+
\frac{1}{2}
\sum_{i,j,s,s'}
V_c(i,j)\,
\hat n_{i,s}\hat n_{j,s'}
+
\sum_q
\omega_q
\left(
\hat d_q^\dagger \hat d_q + \frac{1}{2}
\right),
\end{aligned}
\end{equation}
where $\hat c_{i,s}$ and $\hat d_q$ denote electron and phonon operators, respectively, $\hat n_{i,s}=\hat c_{i,s}^\dagger\hat c_{i,s}$, $t_{i,j}$ is the tight-binding electron tunnelling amplitude, $\omega_q$ is the phonon dispersion relation, $u_{i,s}(q)$ is the electron-phonon coupling matrix element, $\mu$ is the chemical potential, and $V_c(i,j)$ is the Coulomb interaction between electrons.

In the strong electron–phonon coupling regime relevant for 1T-TaS$_2$
\cite{rossnagel2011origin,brazovskii2015modeling,karpov2018modeling,vodeb2019configurational,gerasimenko2019quantum,bozin2023crystallization},
we apply a Lang–Firsov transformation to integrate out the phonons, yielding an effective polaron Hamiltonian~\cite{alexandrov1995polarons}
\begin{equation}
\begin{aligned}
\tilde H
&=
\sum_{i,j,s,s'}
\left(
T_{i,s;j,s'} - \mu \delta_{i,j}
\right)
\delta_{s,s'}
\, \tilde c_{i,s}^\dagger \tilde c_{j,s'}
-
E_p \sum_{i,s} \tilde n_{i,s}
\\
&\quad
+
\frac{1}{2}
\sum_{i,j,s,s'}
v(i,s;j,s')\, \tilde n_{i,s}\tilde n_{j,s'}
+
\sum_q
\omega_q
\left(
\tilde d_q^\dagger \tilde d_q + \frac{1}{2}
\right),
\end{aligned}
\label{eq:H_polaron}
\end{equation}
where the polaron hopping amplitudes $T_{i,s;j,s'}$ are exponentially suppressed compared to $t_{i,j}$, $E_p$ is the polaron binding energy, and $v(i,s;j,s')$ is the effective inter-polaron interaction.
As a result, polarons behave as heavy, weakly mobile particles, and the dominant low-energy physics is governed by charge reorganization rather than coherent transport.

At low temperatures, spin degrees of freedom play no role in the observed dynamics, allowing a reduction to interacting spinless fermions.
Since the kinetic energy of polarons is much smaller than the repulsion, and relaxation proceeds through discrete rearrangements of charge order at domain walls, fermionic exchange statistics are also inessential.
We therefore map the system onto an extended hard-core boson model,
\begin{equation}
\hat H_{\mathrm{HCB}}
=
- t \sum_{\langle i,j\rangle}(\hat b_i^\dagger\hat b_j+\mathrm{h.c.})
+
\frac{1}{2}\sum_{i,j}V_{ij}\hat n_i\hat n_j
-
\mu\sum_i\hat n_i,
\label{eq:H_hcb}
\end{equation}
where $\hat n_i\in\{0,1\}$.
In the strong-coupling regime $V\gg t$, the low-energy spectrum consists of a manifold of metastable charge-ordered configurations separated by energy barriers, consistent with the slow, discrete relaxation observed in STM.

To obtain a minimal effective description directly implementable on analogue quantum hardware, we map this model onto a transverse-field Ising model (TFIM),
\begin{equation}
\hat H_{\mathrm{TFIM}}
=
J\sum_{\langle i,j\rangle}\hat\sigma_i^z\hat\sigma_j^z
-
h_x\sum_i\hat\sigma_i^x
+
h_z\sum_i\hat\sigma_i^z,
\label{eq:TFIM}
\end{equation}
where $\hat\sigma_i^z=\pm1$ encodes local charge occupation. The mapping is derived in the next section and holds only in the weak-$h_x$ limit, where the classical Ising terms favor staggered configurations corresponding to the CDW ground state (Fig.~\ref{fig:1}b), while domain walls between different staggered sectors map directly onto the CDW domain walls observed experimentally (Fig.~\ref{fig:1}c).

Quantum simulations are initialized in metastable domain-wall configurations and evolved via repeated annealing cycles, leading to domain coarsening and relaxation toward the ordered state (Fig.~\ref{fig:1}d).
Although the electronic system supports a larger number of distinct CDW domains due to long-range interactions, the experimentally relevant domain-wall displacement vectors correspond to nearest-neighbor lattice translations, which are faithfully captured by the TFIM.
As a result, the elementary domain-wall reconfiguration processes are identical in experiment and simulation, establishing the TFIM as a minimal yet quantitatively faithful effective model for charge-domain reconfiguration dynamics in 1T-TaS$_2$.

\subsection{Effective model and low-energy mapping}

To identify the microscopic origin of domain-wall motion in the transverse-field Ising model (TFIM), we consider the limit $J\gg h_x$, where domain walls between oppositely ordered staggered regions are well defined.
Starting from Eq.~\ref{eq:TFIM}, we split the Hamiltonian as $H_{\mathrm{TFIM}}=H_0+V$, with $H_0=J\sum_{\langle i,j\rangle}\sigma_i^z\sigma_j^z+h_z\sum_i\sigma_i^z$ and $V=-h_x\sum_i\sigma_i^x$, and derive an effective low-energy description using a Schrieffer--Wolff transformation (see Methods for more details).

\begin{figure}[t]
    \centering
    \includegraphics[width=\linewidth]{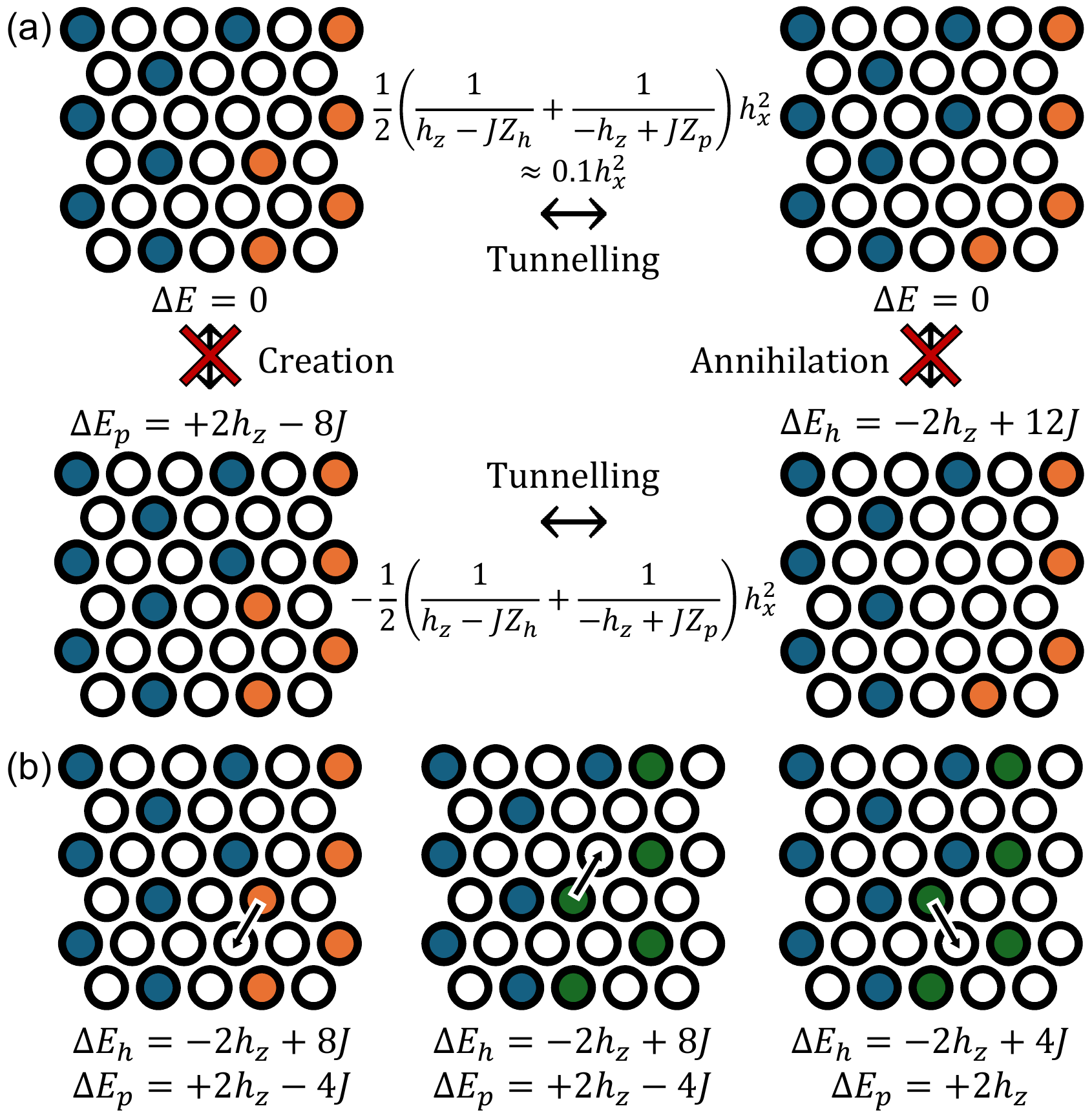}
    \caption{
    (a) Four representative configurations involved in domain wall dynamics in the TFIM. In the effective Hamiltonian obtained via a Schrieffer--Wolff transformation, only second-order tunneling processes survive, connecting states through single polaron tunneling events with matrix elements proportional to $h_x^2$. The associated energy changes $\Delta E$ are shown next to each process, along with the effective tunneling amplitudes. Creation and annihilation of polarons are energetically suppressed, as indicated by red crosses. (b) Additional domain wall types observed in quantum simulation. Black arrows mark the most likely tunnelling events that occur within the domain wall, with intermediate virtual states involving polaron annihilation ($\Delta E_h$) or creation ($\Delta E_p$). Energy costs vary due to local spin environments, illustrating the microscopic diversity of domain wall motion.
    }
    \label{fig:2}
\end{figure}

Our Schrieffer--Wolff analysis does not focus on the full many-body low-energy spectrum of the TFIM, but instead focuses on the experimentally relevant local two-site configurations within a domain wall. Then, the spectrum generically contains two degenerate states related by domain symmetry, while the remaining two states are split off by an energy set by the local environment (Fig.~\ref{fig:2}a).
Eliminating these high-energy virtual states yields the effective Hamiltonian
\begin{equation}
\hat H_{\mathrm{eff}}
=
-h_x^2
\sum_{\langle i,j\rangle}
\lambda_{ij}
\left(
\sigma_i^+\sigma_j^-+\mathrm{h.c.}
\right)
+
\mathcal{O}(h_x^3),
\label{eq:Heff_SW}
\end{equation}
where $\lambda_{ij}=\mathcal{O}(0.1-1)$ depends weakly on $J$ and $h_z$ and on the local domain-wall geometry.
Thus, domain-wall motion arises from second-order tunnelling processes proportional to $h_x^2$, while first-order terms in $h_x$ correspond to single-particle creation or annihilation events.

Identifying the spin states $|{-1}{1}\rangle$ and $|{1}{-1}\rangle$ with occupation states $|01\rangle$ and $|10\rangle$, Eq.~\eqref{eq:Heff_SW} maps directly onto a hard-core boson hopping term with effective amplitude $t\sim h_x^2$.
Crucially, a single-particle hop along a domain wall costs zero classical energy due to translational invariance between neighboring sites.
The same mechanism underlies domain-wall motion in 1T-TaS$_2$, where polarons move within domain walls via direct hopping in the extended hard-core boson model of Eq.~\ref{eq:H_hcb}.
The difference is that, in the TFIM, tunnelling proceeds via virtual intermediate states, leading only to a renormalization of the effective hopping amplitude.

This picture implies that domain-wall relaxation does not occur through coherent collective shifts, but through cascades of local single-particle processes.
Energy relaxation requires first-order particle creation or annihilation events, while second-order tunnelling processes subsequently reposition domain walls.
Such rare-event dynamics are characteristic of strongly interacting polaronic systems, where kinetic energy is strongly renormalized and transport proceeds through localized pathways.

The observed relaxation is further governed by a quasi-static noise environment.
Experimentally, reconfiguration events follow exponential waiting-time statistics and weak temperature dependence, consistent with a stationary Poisson process~\cite{vodeb2024non}.
Microscopically, it is natural to assume that each local emission channel is characterized by a tunnelling rate $\gamma_i=\gamma_0\exp(-B_i)$, where spatial variations of the effective barrier $B_i$ arise from disorder, defects, and the local electrostatic environment.
Summing over many independent channels yields a total rate $\Gamma\propto A$ with suppressed fluctuations.
For typical electronic attempt frequencies in the range $\gamma_0\sim10^{9}$--$10^{12}\,\mathrm{s^{-1}}$, the observed rates imply $B_i\sim10$--$40$, indicating strong tunnelling suppression by quasi-static disorder.

In this regime, noise enters primarily as spatial disorder rather than temporal fluctuations, enabling rare tunnelling events without modifying the local microscopic reconfiguration mechanisms.
An analogous quasi-static noise landscape is present in the quantum simulations, where device noise plays the same role, allowing direct comparison between experimental and simulated domain-wall dynamics, explored in a subsequent section.

\subsection{Quantum simulation protocol and Hamiltonian parametrization}

The domain-wall dynamics are simulated on a programmable superconducting quantum annealer (D-Wave Advantage6.1), which realizes an open quantum system governed by a time-dependent transverse-field Ising Hamiltonian
\begin{equation}
\hat H_{\mathrm{DW}}(s)
=
-\frac{A(s)}{2}\sum_i \hat\sigma_i^x
+
\frac{B(s)}{2}
\left(
\sum_{i,j} J_{ij}\hat\sigma_i^z\hat\sigma_j^z
+
\sum_i h_i^z \hat\sigma_i^z
\right),
\label{eq:H_DW_AB}
\end{equation}
where $s\in[0,1]$ is the annealing schedule parameter and $A(s)$ and $B(s)$ control the transverse-field and longitudinal Ising energy scales, respectively.
The Pauli operators act on logical spins encoding binary charge degrees of freedom.

To connect this programmable Hamiltonian to the effective domain-wall model and to characterize noise effects, we rewrite it in dimensionless form.
We decompose the Ising couplings as $J_{ij}=J\,\tilde J_{ij}$, where $\tilde J_{ij}$ defines the interaction geometry and $J$ is an overall tunable scale by the user.
We factor out $rA(s)/2$ and parametrize $J=\alpha r$ with $\alpha=A(s)/B(s)$ and $r$ a dimensionless model parameter.
With this choice, the leftover $B(s)/A(s)$ ratio cancels with $\alpha$ during each annealing schedule, yielding
\begin{equation}
\begin{aligned}
\frac{\hat H_{\mathrm{DW}}}{k_\mathrm{B}T}
&=
\frac{A(s)r}{2k_\mathrm{B}T}
\bigg[
-\frac{1}{r}\sum_i \hat\sigma_i^x
\\
&\quad
+
\frac{B(s)}{A(s)}\,\alpha
\left(
\sum_{i,j} \tilde J_{ij}\hat\sigma_i^z\hat\sigma_j^z
+
\sum_i \tilde h_i^z \hat\sigma_i^z
\right)
\bigg].
\end{aligned}
\label{eq:H_DW_Teff}
\end{equation}
where the effective temperature is defined as
$T_\mathrm{eff}[s(t)]=2k_\mathrm{B}T/A(s)r$.
In this representation, $r$ controls the ratio between interaction energy and quantum tunnelling, while $T_\mathrm{eff}$ captures the strength of noise relative to the Hamiltonian energy scales.
Both parameters and the annealing schedule are independently tunable by the user.

The reduced couplings are chosen as $\tilde J_{ij}=1$ for nearest neighbors on the underlying triangular lattice and zero otherwise, implementing antiferromagnetic interactions.
The reduced longitudinal fields $\tilde h_i^z$ encode the chemical potential for charge occupation and are set to $4$ for bulk sites with six nearest neighbors, decreasing to $3,2,1$, or $0$ for sites with reduced coordination.

\begin{figure}[t]
    \centering
    \includegraphics[width=\linewidth]{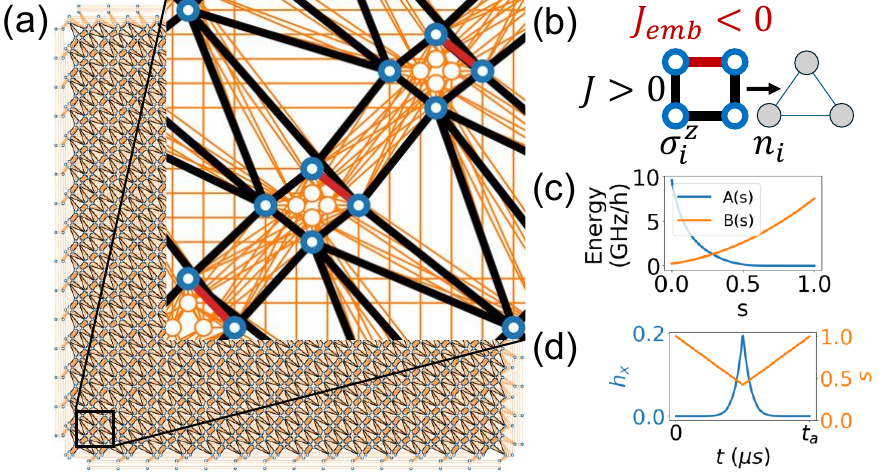}
    \caption{
    \textbf{Quantum simulation of domain-wall dynamics on a superconducting quantum annealer.}
    (a)~Embedding of a triangular lattice with $N_\mathrm{L}=2008$ logical sites into the quantum annealer using $N_\mathrm{P}=2673$ physical qubits.
    Each square plaquette of four physical qubits represents a building block triangle of the logical lattice, resulting in a physical-to-logical qubit ratio of $4/3$.
    Inset: detailed view of the embedding.
    (b)~Schematic of the embedding couplings.
    Black links denote antiferromagnetic Ising couplings ($J>0$) that map onto the edges of the logical triangular lattice.
    Red links indicate strong ferromagnetic embedding couplings ($J_\mathrm{emb}<0$) that bind pairs of physical qubits into a single logical qubit by enforcing identical spin states.
    (c)~Energy scales of the quantum annealer Hamiltonian as functions of the annealing schedule parameter $s$.
    The transverse-field term $A(s)$ and the Ising term $B(s)$ are shown in units of GHz$/h$, where $h$ is Planck’s constant.
    (d)~Reverse annealing protocol used to generate relaxation dynamics.
    Starting from a classical configuration at $s=1$ ($h_x=0$), the transverse field is ramped to a target value corresponding to the desired $h_x=1/r$ over an annealing time $t_a=5~\mu s$ and then ramped back to $s=1$.
    The final configuration of each reverse anneal is used as the initial configuration for the next cycle, producing a discrete-time dynamical evolution analogous to time-resolved STM measurements.
    }
    \label{fig:3}
\end{figure}

The logical triangular lattice consists of $N_\mathrm{L}=2008$ sites and is embedded into the hardware graph using standard minor-embedding techniques, requiring $N_\mathrm{P}=2673$ physical qubits (physical-to-logical ratio $4/3$).
Each elementary triangle is represented by four physical qubits, with antiferromagnetic couplers implementing logical interactions and strong ferromagnetic couplings ($-2J$) binding pairs of physical qubits into a single logical spin (Fig.~\ref{fig:3}a,b).

Relaxation dynamics are generated using a reverse annealing protocol.
Each cycle starts from a classical configuration at $s=1$ (vanishing transverse field), the schedule is reversed to a target value of $s$ over an annealing time $t_a=5~\mu$s to enable tunnelling, and the system is then returned to $s=1$ for readout. The energy scales $A(s)$ and $B(s)$ and the reverse annealing schedule are shown in Fig.~\ref{fig:3}c,d.
The final configuration of each cycle is used as the initial condition for the next, producing a discrete-time dynamical evolution.
The iteration index thus defines an effective time variable directly analogous to the sequence of time-resolved STM snapshots.

\subsection{Scaling collapse of reconfiguration rates and noise-dominated dynamics}

We quantify the domain-wall reconfiguration dynamics by the average fraction of spins that flip between successive reverse-annealing iterations,
\begin{equation}
R
=
\frac{1}{N_{\mathrm{steps}}}
\sum_{t=1}^{N_{\mathrm{steps}}}
\frac{1}{2N}
\sum_{i=1}^{N}
\left|
\sigma_i^z(t+1)-\sigma_i^z(t)
\right|,
\end{equation}
which directly measures domain-wall activity and is the discrete-time analogue of the experimentally extracted polaron reconfiguration rate.

\begin{figure}[t]
    \centering
    \includegraphics[width=\linewidth]{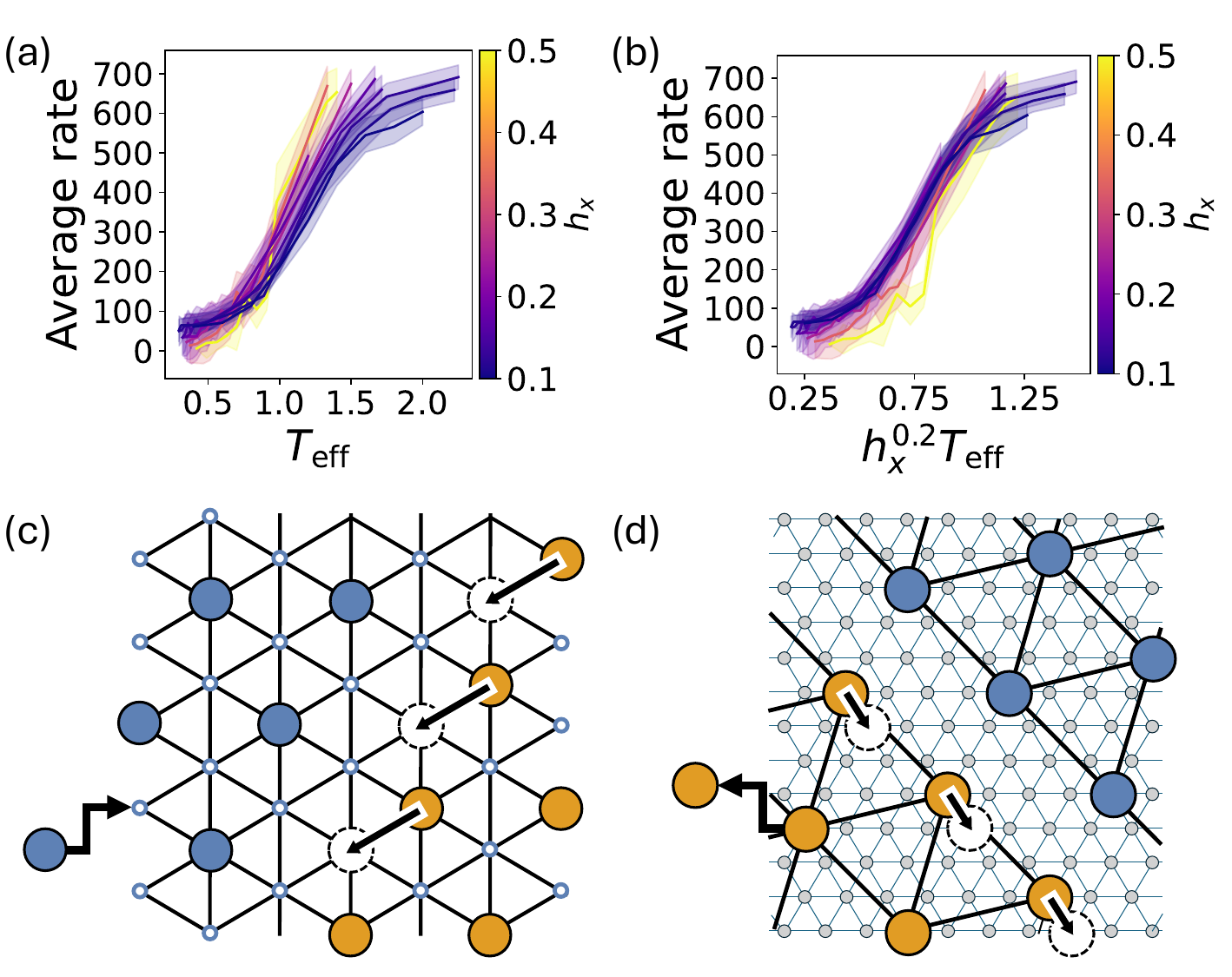}
    \caption{
    \textbf{Scaling collapse of reconfiguration rates and noise-dominated dynamics.}
    (a)~Average reconfiguration rate as a function of effective temperature $T_\mathrm{eff}$ for several transverse-field strengths $h_x$, showing a smooth crossover from weakly temperature-dependent dynamics at low $T_\mathrm{eff}$ to rapidly increasing activity at higher $T_\mathrm{eff}$.
    The dependence on $h_x$ is comparatively weak, with only a modest shift of the crossover.
    (b)~Scaling collapse of the same data obtained by rescaling the temperature axis as $T_\mathrm{eff}\rightarrow h_x^{0.2}T_\mathrm{eff}$.
    The near-collapse across more than an order of magnitude in rate indicates that the reconfiguration dynamics depend only weakly on $h_x$, inconsistent with a tunnelling-dominated scaling $\propto h_x^2$ or higher order processes.
    (c)~Schematic of elementary reconfiguration processes on the effective triangular lattice.
    Large apparent domain-wall rearrangements arise from cascades of local single-polaron tunnelling events (dashed arrows) rather than collective multi-polaron tunnelling.
    (d)~Corresponding schematic for the experimental system, illustrating that extended rearrangements of CDW domain walls are composed of sequences of local polaron moves within the domain-wall network.
    Despite differences in global charge evolution between simulation and experiment, the microscopic reconfiguration mechanisms at domain walls are the same in both systems.
    }
    \label{fig:4}
\end{figure}

Figure~\ref{fig:4}a shows $R$ as a function of the effective temperature $T_\mathrm{eff}$ for several transverse-field strengths $h_x$.
All curves exhibit a smooth crossover from weakly temperature-dependent dynamics at low $T_\mathrm{eff}$ to rapidly increasing activity at higher $T_\mathrm{eff}$.
The dependence on $h_x$ is, however, remarkably weak, with only a modest shift of the crossover.

In the weak-$h_x$ regime, the effective domain-wall Hamiltonian can be expanded by first factoring out the dominant interaction scale $r=1/h_x\gg1$.
This procedure yields a low-energy generator for configurational changes whose leading tunnelling contribution scales as $1/r^2\propto h_x^2$, in direct analogy to the Schrieffer--Wolff construction of the low-temperature effective model.
If the dynamics were tunnelling dominated, a strong $h_x^2$ dependence of $R$ would therefore be expected.

To test this, we perform a scaling collapse by rescaling the temperature axis as $T_\mathrm{eff}\rightarrow h_x^{n}T_\mathrm{eff}$.
As shown in Fig.~\ref{fig:4}b, the best collapse is obtained for $n\simeq0.2$, indicating that the effective scale governing the dynamics depends only weakly on $h_x$ and is incompatible with a tunnelling-controlled $h_x^2$ scaling.

This behavior points to a noise-dominated regime.
On the quantum annealer, coherent tunnelling processes generated by the transverse field occur on intrinsic nanosecond timescales set by the inverse transverse-field energy.
In contrast, the observed reconfiguration dynamics unfold on microsecond timescales determined by the reverse-annealing protocol.
This large separation of timescales implies that tunnelling itself is not rate limiting.
Rather, the dominant bottleneck is the slow, quasi-static evolution of local biases and detunings that intermittently enable tunnelling events~\cite{whiticar2023probing}.

This can be made explicit by considering a minimal two-site model representing a local tunnelling process inside a domain wall.
In the basis $\{|10\rangle,|01\rangle\}$, corresponding to a single polaron occupying one of two neighboring sites, the effective Hamiltonian takes the form
$H = \frac{\Delta}{2}\sigma^z + t\,\sigma^x$,
where $t$ is the tunnelling matrix element (set by $t\sim h_x^2$ in the effective model) and $\Delta$ is the energy detuning between the two sites.
In the absence of detuning ($\Delta=0$), the system undergoes coherent Rabi oscillations with frequency $\Omega=2t$, and a tunnelling event occurs on a timescale $\tau\sim 1/t$, which is on the order of nanoseconds for the parameters of the quantum annealer.

In the presence of a quasi-static detuning $\Delta$, however, the oscillation frequency becomes $\Omega=\sqrt{(2t)^2+\Delta^2}$ and, more importantly, the tunnelling probability is suppressed by a factor $(2t/\Omega)^2$.
For $|\Delta|\gg t$, tunnelling is effectively frozen, and transitions occur only when slow noise fluctuations reduce $|\Delta|$ to values comparable to or smaller than $t$.
Thus, tunnelling acts as a fast process that is conditionally activated when the quasi-static noise brings two sites into near resonance.
The observed reconfiguration rate is therefore set not by $t$ itself, but by the rate at which the noise landscape produces such resonant conditions.

This simple estimate also provides a quantitative constraint on the noise amplitude.
Given that the intrinsic tunnelling scale corresponds to nanosecond dynamics while reconfiguration events are observed on microsecond timescales, the typical detuning induced by quasi-static noise must exceed the tunnelling matrix element by at least one order of magnitude.
Only rare fluctuations that transiently reduce the detuning below this scale allow a tunnelling event to occur.
Once activated, the tunnelling process completes rapidly, explaining the weak dependence of the reconfiguration rate on the transverse field $h_x$.

Although snapshots reveal large apparent domain-wall rearrangements, these arise from cascades of such local single-polaron tunnelling events rather than collective multi-polaron motion (Fig.~\ref{fig:4}c,d).
The same mechanism operates in experiment and simulation: strong local interactions ensure rapid internal rearrangement once a reconfiguration is triggered, while the waiting time between events is governed by the quasi-static noise landscape.
The scaling collapse therefore provides direct evidence that relaxation dynamics are controlled primarily by noise-assisted resonance conditions, rather than by the intrinsic tunnelling amplitude itself.

\section{Discussion}

In the broader context of quantum simulation, this work demonstrates a complementary paradigm to existing approaches.
While quantum simulators have successfully been used to study molecular chemistry~\cite{kandala2017hardware}, lattice gauge theories~\cite{dalmonte2016lattice}, disorder-driven localization~\cite{schreiber2015observation}, programmable quantum matter~\cite{ebadi2021quantum,hirthe2023magnetically}, and quantum phase transitions~\cite{greiner2002quantum,king2023quantum}, these studies typically focus on idealized models with prospective relevance to materials.
Here, we instead use quantum simulation as a microscope for a concrete, experimentally accessible solid-state system, directly linking programmable quantum dynamics to microscopic non-equilibrium relaxation processes in a strongly correlated material.

By combining effective low-energy modeling, large-scale quantum simulations, and dynamical scaling analysis, we show that the slow relaxation of metastable domain-wall networks in 1T-TaS$_2$ is governed by cascades of local, noise-assisted tunnelling events rather than coherent collective motion.
The quantum annealer enables controlled access to regimes that are difficult to isolate experimentally, allowing us to disentangle intrinsic tunnelling processes from the role of quasi-static noise.
This establishes quantum annealing as a practical tool for identifying microscopic relaxation pathways in complex materials, rather than merely reproducing equilibrium properties.

The relevance of this approach extends well beyond 1T-TaS$_2$.
Metastable states and their decay pathways play a central role in a wide range of correlated electron systems, including resistive switching oxides~\cite{jeong2013suppression,wang2018threshold}, electronically phase-separated manganites~\cite{miyano1997photoinduced,zhang2002direct}, photo-induced superconductivity~\cite{budden2021evidence}, and other charge-density-wave materials~\cite{stojchevska2014ultrafast,vaskivskyi2015controlling}.
In many such systems, macroscopic functionality is controlled by microscopic relaxation mechanisms that remain experimentally opaque.
Quantum simulations capable of distinguishing collective tunnelling from cascaded local dynamics, as demonstrated here, may therefore provide valuable guidance for material design strategies aimed at stabilizing or exploiting metastable states.

An important experimental implication of our results concerns the direct detection of the quasi-static noise landscape that inhibits polaron tunnelling.
Using scanning tunnelling spectroscopy (STS), it should be possible to probe the local electronic spectrum of a single site within a domain over timescales comparable to the measured reconfiguration rate.
Temporal fluctuations of the local spectrum would provide direct evidence for noise-induced modulation of tunnelling barriers.
In particular, correlated shifts of the local density of states on pairs of sites involved in a tunnelling event would strongly support a noise-assisted mechanism: when the local densities of states align, a tunnelling event should occur.
Conversely, persistent spectral mismatches between candidate tunnelling sites would suppress reconfiguration.
To be detectable, such spectral shifts must exceed the intrinsic polaron bandwidth of approximately $80\,\mathrm{meV}$ in 1T-TaS$_2$.
Time-resolved STS measurements thus offer a concrete route to directly observe the noise processes inferred from our analysis.

Looking ahead, applying similar quantum-simulation-based analyses to experiments with higher temporal resolution could further resolve the internal structure of tunnelling cascades.
On the simulation side, extending the effective model beyond the transverse-field Ising description to include fermions, explicit phonon coupling, disorder, or longer-range interactions would provide deeper insight into how environmental factors shape noise-assisted relaxation dynamics.
Together, these directions point toward a general framework in which quantum simulation and local spectroscopy jointly uncover the microscopic origins of non-equilibrium behavior in strongly correlated quantum materials.


\section{Methods}

The Hamiltonian after the SW transformation is $H^{'}=e^SHe^{-S}=H_0+\frac{1}{2}[S,V]+O(V^3)$, where $S$ is found through solving $V+[S,H_0]=0$. We limit ourselves to a two-spin subspace that exists within an already existing domain wall, shaped as the most typical observed domain walls found in 1T-TaS$_2$ and quantum simulation. The four possible configurations within this subspace are shown in Fig.~\ref{fig:2}a. Within this subspace, we write 
\[
H_0 = \sum_{\psi,\psi'} \langle \psi | H_0 | \psi' \rangle \, |\psi\rangle \langle\psi'|, \quad \text{where } 
\]
\[
|\psi\rangle \in \{|{-1}{-1}\rangle, |{-1}{1}\rangle, |{1}{-1}\rangle, |{1}{1}\rangle\}
\]
\[
H_0 = \bigg(\begin{smallmatrix}
  -2 h_z + 2 J Z_h & 0 & 0 & 0 \\
  0 & 0 & 0 & 0 \\
  0 & 0 & 0 & 0 \\
  0 & 0 & 0 & 2 h_z - 2 J Z_p
\end{smallmatrix}\bigg)
\]
\[
V = \bigg(\begin{smallmatrix}
0 & -h_x & -h_x & 0 \\
-h_x & 0 & 0 & -h_x \\
-h_x & 0 & 0 & -h_x \\
0 & -h_x & -h_x & 0
\end{smallmatrix}\bigg).
\]
The energy of the initial domain state in the top left of Fig.~\ref{fig:2}a is subtracted from $H_0$ as the reference energy. Using them, we find
\[
\alpha = \frac{h_x}{2(h_z - J Z_h)}, \quad
\beta = \frac{h_x}{2(h_z - J Z_p)}
\]
\[
S = \bigg(\begin{smallmatrix}
0 & \alpha & \alpha & 0 \\
-\alpha & 0 & 0 & \beta \\
-\alpha & 0 & 0 & \beta \\
0 & -\beta & -\beta & 0
\end{smallmatrix}\bigg),
\]
which we then use to calculate 
\[
\lambda = \frac{1}{2} \left( \frac{1}{h_z - J Z_h} + \frac{1}{-h_z + J Z_p} \right)
\]

\[
H' = \lambda h_x^2
\begin{pmatrix}
\displaystyle \frac{2(-h_z + J Z_p)}{J(Z_h - Z_p)} & 0 & 0 & -1 \\
0 & 1 & 1 & 0 \\
0 & 1 & 1 & 0 \\
-1 & 0 & 0 & \displaystyle \frac{2(h_z - J Z_h)}{J(Z_h - Z_p)}
\end{pmatrix}
\]

Crucially, first-order terms $\sim h_x$ corresponding to single spin flips are energetically forbidden (matrix elements equal to $0$) within domain wall regions, as flipping a spin in the antiferromagnetic background changes the energy by either $-2h_z + 2JZ_h$ or $2h_z - 2JZ_h$. As a result, the leading contributions arise from second-order processes in $h_x$.

These second-order terms generate effective hopping between domain wall configurations. The resulting effective Hamiltonian within the subspace $|{-1}{1}\rangle, |{1}{-1}\rangle$ takes the form
\[
    \begin{aligned}
        \hat{H}_\mathrm{eff} &= -h_x^2\sum_{\langle i,j \rangle} (\lambda_{i,j}\hat{\sigma}_i^+ \hat{\sigma}^-_j + \mathrm{h.c.}) + \mathcal{O}(h_x^3),
    \end{aligned}
\]
where $\hat{\sigma}_i^+$ and $\hat{\sigma}_i^-$ are the raising and lowering operators for spins. Here, we intentionally make the change in notation $\lambda\rightarrow\lambda_{i,j}$ to emphasize the dependence of $\lambda_{i,j}$ on the choice of the local environment surrounding sites $i$ and $j$ in the 2-site model. If we reinterpret the states $|{-1}{1}\rangle, |{1}{-1}\rangle$ as occupation operator states $|{0}{1}\rangle, |{1}{0}\rangle$, we can rewrite $\hat{H}_{eff}$ using $b_i^\dagger$ and $b_i$ as the hardcore boson creation and annihilation operators, into the standard hopping term with $t \sim h_x^2$ as the effective hopping amplitude.


\section{Data Availability}

Quantum simulation data obtained from the quantum annealer were deposited to Zenodo and are available at the following URL: \url{https://doi.org/10.5281/zenodo.18089475}.\\


\section{Acknowledgments}

We thank Dragan Mihailović and Denis Golež for the fruitful discussions.


\section{Funding}

This study was funded by the Slovenian Research Agency grants P1-0040 and P1-0416 as well as ERC AdG grant ‘HIMMS’. The funder played no role in study design, data collection, analysis and interpretation of data, or the writing of this manuscript. 


\section{Author Contributions}

J.V. developed the idea, performed and analyzed the quantum simulations, derived the effective Hamiltonian, and wrote the manuscript.


\section{Competing interests}

The authors declare no competing interests.


\bibliographystyle{apsrev4-2}
\bibliography{references}


\end{document}